\documentclass[twocolumn]{aa}
\usepackage{graphicx}
\usepackage{lscape}

\begin{document}
   \title{High Frequency Peakers:\\ young radio sources or flaring blazars?}

      \author{S. Tinti
          \inst{1} \and D. Dallacasa\inst{2,3} \and G. De Zotti \inst{4,1}
          \and A. Celotti \inst{1} \and C. Stanghellini\inst{3,5}}

   \offprints{S. Tinti,\email{tinti@sissa.it}}

\institute{SISSA/ISAS, Via Beirut 4, 34014, Trieste, Italy  \and
Dipartimento di Astronomia, Via Ranzani 1, I-40127, Bologna, Italy
\and Istituto di Radioastronomia - CNR, Via Gobetti 101, I-40129
Bologna, Italy  \and INAF, Osservatorio Astronomico, Vicolo
dell'Osservatorio 5, 35122, Padova, Italy  \and Istituto di
Radioastronomia - CNR, C. P. 169, I-96017 Noto (SR), Italy}

   \date{...}

   \abstract{We present new, simultaneous, multifrequency observations
   of 45 out of the 55 candidate High Frequency Peakers (HFP) selected by
   Dallacasa et al. (2000), carried out 3 to 4 years after a first set
   of observations. Our sub-sample consists of 10 galaxies, 28 stellar objects
   (``quasars'') and 7 unidentified sources. Both sets of observations are
   sensitive enough to allow the detection of variability at the 10\% level
   or lower. While galaxies do not show significant variability, most quasars
   do. Seven of them no longer show the convex spectrum which is the defining
   property of Gigahertz Peaked Spectrum (GPS)/HFP sources and are interpreted 
   as blazars caught by Dallacasa et al. (2000) during a flare, when a highly 
   self-absorbed component dominated the emission.
   In general, the variability properties (amplitude, timescales, correlation
   between peak luminosity and peak frequency of the flaring component) of the
   quasar sub-sample resemble those of blazars. We thus conclude that most HFP
   candidates identified with quasars may well be flaring blazars.

   }

   \maketitle
%

\section{Introduction}

The Compact Steep Spectrum (CSS) and Gigahertz Peaked Spectrum
sources (GPS) are two classes of intrinsically compact objects
(linear size $<10$--1 kpc) defined on the basis of their spectral
properties: the overall shape is convex with turnover frequencies
between 100 MHz (CSS) and a few GHz (GPS) and the spectral index at
high frequencies is steep (see O'Dea 1998 for a complete review).

The currently accepted model (the {\textit{youth scenario}})
relates the small linear size of GPS/CSS sources to their age,
implying that they are the progenitors of extended radio sources.
Both kinematic studies (Owsianik \& Conway 1998; Owsianik,
Conway \& Polatidis 1998) and spectral analysis (Murgia et al.
1999) indicate ages of $10^3$--$10^5$ yr. On the other hand,
various observations across the electromagnetic spectrum (e.g.
Fanti et al. 2000) failed to reveal the presence of a particularly
dense medium that could prevent the expansion of the radio lobes
in the  intergalactic medium, as proposed in the
{\textit{frustration model}} (van Breugel 1984; Baum et al. 1990).
So far no evidence of differences in the properties of the host
galaxies between GPS/CSS and FRII radio sources has been reported
(e.g. see Fanti et al. 2000).

Observational studies of the population of GPS/CSS sources have
led to the discovery  of an anti-correlation between the radio
turnover frequency, $\nu_p$, and the projected angular size,
$\theta$ (O'Dea \& Baum 1997; Fanti et al. 2002). This
relationship is expected if the peaked component is due to 
synchrotron self-absorption, where
$\theta^2\propto \nu_p^{-5/2}$ , although free-free absorption
could also play a role (Bicknell, Dopita \& O'Dea 1997). In the
youth scenario this correlation means that the youngest objects
have the highest turnover frequencies and that the peak frequency
is expected to move towards lower frequencies as the source
expands and the energy density decreases. In principle, the
highest the turnover frequency, the youngest the radio source is.

The shape of the radio spectrum and the position of the turnover
have been used as selection tools for this class of objects.
Dallacasa et al. (2000) selected a sample of ``bright'' radio
sources with turnover frequencies above 5 GHz and called them High
Frequency Peakers (HFPs). An early summary of their properties can
be found in Dallacasa (2003). Simultaneous multifrequency VLA
observations were made to define the spectral shape of all
candidates. The comparison with literature data showed that flux
density variability is not uncommon among the candidates and that
also beamed radio sources like blazars can meet the selection
criteria of the sample, e.g. when a flaring, strongly
self-absorbed synchrotron component dominates the emission
spectrum. A substantial contamination of the HFP sample by blazars
may obviously lead astray analyses of HFP properties.

In this paper we present the results of a second set of
simultaneous multifrequency VLA observations of the sample of
bright HFPs, aimed at discriminating between bona-fide HFPs and
blazars, and at studying the spectral evolution of the
sources. At 1.465 and 1.665 GHz, longer observations were
carried out to search for extended emission around the point-like
dominant component. It is known that a significant fraction ($\sim
10\%$) of GPS sources do have extended emission (e.g. Stanghellini
2003). Also polarization
data have been taken and will be presented elsewhere.

This paper is structured in the following way: Section 2 describes
the radio properties of the sample, the second-epoch VLA
observations and the data reduction; Section 3 presents the
analysis of simultaneous radio spectra; Section 4 deals with
the extended emission; in Section 5 the variability properties are
investigated; the main conclusions are summarized and discussed in
Section 6. Throughout this paper, we adopt $H_0=50 km s^{-1} Mpc^{-1}$
and $\Omega=1$

   \begin{figure}
   \centering
   \includegraphics[width=8 cm]{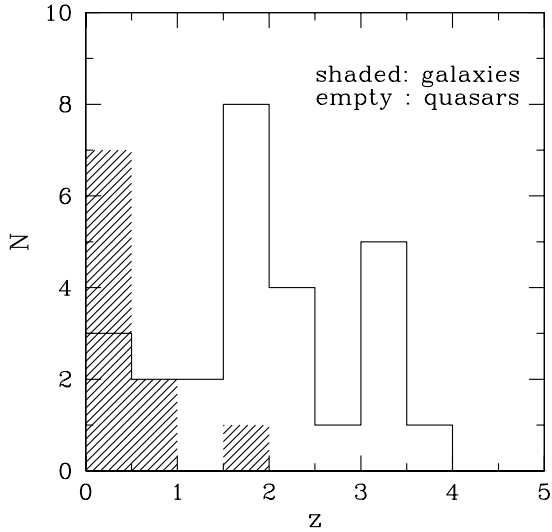}
      \caption{Distribution of measured redshifts for
      the sample by Dallacasa et al. (2000).}
      \label{fig:dist_redshift}
   \end{figure}

\section{The Bright HFP Sample}

\subsection{Selection and radio properties of the sample}

The bright HFP sample of Dallacasa et al. (2000) was selected by
cross-correlating the 87GB (Gregory et al. 1996) sources with
$S_{4.9{\rm GHz}} \geq 300\,$mJy with the NVSS catalogue (Condon
et al. 1998) at 1.4 GHz and picking out those with inverted
spectra ($\alpha < -0.5$, $S\propto\nu^{-\alpha}$). Since the 87GB
and the NVSS surveys were carried out at different epochs, this
sample is contaminated by flat-spectrum sources that happened to
be flaring when they were observed at 4.9 GHz. The sample was then
``cleaned'' by means of simultaneous multifrequency VLA
observations at 1.365, 1.665, 4.535, 4.985, 8.085, 8.485, 14.96
and 22.46 GHz, leaving 55 sources whose single-epoch radio
spectrum peaks at frequencies ranging from a few GHz to about 22
GHz. These ``first epoch'' observations were carried out in
1998--1999.

The final sample of HFP candidates comprises 11 galaxies
(including a type 1 Seyfert), 36 quasars, and 8 still unidentified
sources (Dallacasa et al. 2002a). For comparison, galaxies and
quasars are almost equally represented in GPS samples
(Stanghellini et al. 1998) and most CSS sources are galaxies
(Dallacasa 2002b,c). 36 objects out of 55 have known redshift and
their distribution (Fig.~\ref{fig:dist_redshift}) confirms the
trend of quasars and galaxies to have barely overlapping redshift
ranges. The redshift distribution of quasars shows a much slower
decline with increasing redshift at $z>2$ than flat-spectrum
quasars in the 2.7 GHz Parkes quarter Jansky sample (Jackson et
al. 2002) with a similar flux-density limit (although at a lower
frequency). This difference can be due, at least partly, to the
more favorable K-correction for high-$z$ HFPs associated with
their inverted spectra, and to the increasing variability
amplitude with increasing frequency (Impey \& Neugebauer 1988;
Ciaramella et al. 2004), which also enhances the visibility of
distant quasars.

\begin{table*}
\begin{center}
\caption{Integrated flux densities of candidates HFP sources
at the 2002 epoch.}
\begin{tabular}{cllr||crrrrrrrr}
\hline
\hline
J2000&ID &\multicolumn{1}{c}{$z$}&  $S_{5.0}^{\rm old}$ & &$S_{1.4}$&$S_{1.7}$&$S_{4.5}$&$S_{4.9}$&$S_{8.1}$&$S_{8.5}$& $S_{15.0}$&$S_{22.5}$  \\
Name  &   &    &               mJy& & mJy  & mJy & mJy & mJy & mJy & mJy & mJy& mJy \\
\hline
0003+2129 &  G* & 0.4    &  265 &  C &102 & 120 & 250 & 253 & 234 & 227 & 140 & 86  \\
0005+0524 &  Q  & 1.887  &  229 &  C &168 & 186 & 220 & 213 & 171 & 166 & 111 & 82  \\
0037+0808 &  G? &  $>$1.8&  292 &  C &101 & 118 & 281 & 283 & 267 & 262 & 190 & 143 \\
0111+3906 &  G  & 0.668  & 1324 &  E &476 & 594 &1313 &1286 & 978 & 937 & 507 & 315 \\
0116+2422 &     &        &  243 &  C &110 & 124 & 254 & 258 & 252 & 248 &     & 149 \\
0217+0144 &  Q  & 1.715  &  1862&  E &810 & 802 & 743 & 745 & 810 & 810 & 856 & 838 \\
0329+3510 &  Q* &        &  770 &  E &442 & 462 & 560 & 555 & 608 & 613 & 659 & 702 \\
0357+2319 &  Q* &        &  560 &  C &139 & 133 & 131 & 130 & 143 & 144 & 154 & 168 \\
0428+3259 &  G* & 0.3    &  506 &  E &177 & 202 & 493 & 506 & 525 & 514 & 375 & 263 \\
0519+0848 &     &        &  278 &  E &279 & 296 & 446 & 446 & 435 & 430 & 420 & 401 \\
0625+4440 & BL  &        &  442 &  C &172 & 184 & 231 & 231 & 238 & 238 & 219 & 210 \\
0638+5933 &     &        &  591 &  C &277 & 301 & 590 & 606 & 668 & 667 & 620 & 567 \\
0642+6758 &  Q  & 3.180  &  474 &  C &245 & 288 & 429 & 417 & 331 & 321 & 203 & 149 \\
0646+4451 &  Q  & 3.396  & 1896 &  C &522 & 612 &2580 &2770 &3709 &3757 &3691 &3318 \\
0650+6001 &  Q  & 0.455  & 1236 &  C &517 & 625 &1142 &1136 & 989 & 964 & 671 & 495 \\
1335+4542 &  Q  & 2.449  &  735 &  C &294 & 366 & 797 & 785 & 613 & 592 & 359 & 234 \\
1335+5844 &     &        &  723 &  C &319 & 396 & 734 & 725 & 680 & 671 & 531 & 253 \\
1407+2827 &  G  & 0.0769 & 2362 &  C &910 &1100 &2469 &2463 &2114 &2050 &1139 & 604 \\
1412+1334 &     &        &  330 &  C &205 & 236 & 345 & 337 & 275 & 267 & 185 & 130 \\
1424+2256 &  Q  & 3.626  &  607 &  C &352 & 414 & 686 & 669 & 479 & 460 & 251 & 145 \\
1430+1043 &  Q  & 1.710  &  910 &  C &326 & 402 & 885 & 882 & 770 & 752 & 546 & 385 \\
1457+0749 &     &        &  241 &  E &360 & 367 & 262 & 251 & 194 & 188 & 141 & 109 \\
1505+0326 &  Q  & 0.411  &  929 &  C &413 & 465 & 740 & 741 & 721 & 710 & 665 & 567 \\
1511+0518 & Sy1 & 0.084  &  536 &  C &90  & 115 & 568 & 607 & 848 & 861 & 843 & 617 \\
1526+6650 &  Q  & 3.02   &  411 &  C &115 & 145 & 408 & 410 & 352 & 341 & 193 & 107 \\
1603+1105 &     &        &  270 &  C?&182 & 188 & 214 & 214 & 225 & 225 & 234 & 217 \\
1616+0459 &  Q  & 3.197  &  892 &  C &317 & 382 & 787 & 771 & 582 & 559 & 333 & 212 \\
1623+6624 &  G  & 0.203  &  298 &  C &159 & 175 & 295 & 298 & 287 & 283 & 224 & 175 \\
1645+6330 &  Q  & 2.379  &  513 &  C &303 & 315 & 510 & 526 & 615 & 618 & 596 & 493 \\
1735+5049 & G?  &        &  968 &  C &448 &     & 925 & 943 & 934 & 920 & 740 & 587 \\
1800+3848 &  Q  & 2.092  &  791 &  E &313 & 331 & 702 & 748 & 1044& 1063&1174 & 1076\\
1811+1704 &  Q* &        &  691 &  E &545 & 530 & 494 & 491 & 508 & 509 & 499 & 418 \\
1840+3900 &  Q  & 3.095  &  203 &  C &125 & 130 & 169 & 167 & 161 & 158 & 134 & 114 \\
1850+2825 &  Q  & 2.560  & 1246 &  C &203 & 223 &1096 &1185 &1540 & 1541& 1318&1045 \\
1855+3742 &  G* & 0.5    &  364 &  C &193 & 191 & 362 & 345 & 222 & 212 & 124 & 91  \\
2021+0515 &  Q* &        &  477 &  C &260 & 302 & 450 & 441 & 377 & 368 & 267 & 191 \\
2024+1718 &  S  & 1.050? &  572 &  E &280 & 304 & 676 & 704 & 803 & 800 & 697 & 569 \\
2101+0341 &  Q  & 1.013  &  920 &  E &553 & 610 & 694 & 678 & 593 & 583 & 499 & 478 \\
2123+0535 &  Q  & 1.878  & 1896 &  E &1954&2006 &2084 &2111 &2462 &2482 &2755 &2560 \\
2203+1007 &  G* & 0.9    &  319 &  C &124 & 123 & 312 & 306 & 243 & 234 & 129 & 77  \\
2207+1652 &  Q* &        &  551 &  E &223 &     & 249 & 247 & 227 & 223 & 188 & 163 \\
2212+2355 &  S  &        & 1182 &  C &601 &     & 963 & 979 &1033 &1028 & 975 & 915 \\
2257+0243 &  Q  & 2.081  &  274 &  C &174 &     & 274 & 292 & 439 & 450 & 558 & 528 \\
2320+0513 &  Q  & 0.622  & 1196 &  E &640 &     & 657 & 656 & 720 & 725 & 806 & 843 \\
2330+3348 &  Q  & 1.809  &  558 &  C &370 & 383 & 407 & 407 & 457 & 463 & 532 & 548 \\
\hline
\hline
\end{tabular}
\label{tab:VLA02_flux}
\end{center}
{}$^\ast$ indicates the sources optically identified by Dallacasa
et al. (2002a).
\end{table*}

\subsection{Observations and data reduction}

In July 2002 we carried out multifrequency radio observations of
the 45 (out of 55) candidate HFPs that were visible
during the allocated observing time. We used the VLA in the B
configuration, with a frequency coverage similar to that used for
the sample definition. This sub-sample comprises 10 galaxies
(including the type 1 Seyfert), 28 stellar objects, henceforth
referred to as ``quasars'' (including a BL Lac), and 7
unidentified sources. We used the intermediate frequencies (IFs)
of 1.465 and 1.665 GHz in the L band, of 4.565 and 4.935 GHz in
the C band, and of 8.085 and 8.465 GHz in the X band, while the
standard VLA frequencies were used in the U band (14.96 GHz) and
in the K band (22.46 GHz). The observing bandwidth was chosen to
be 50 MHz per IF. Separate analysis for each IF in L, C and X
bands was carried out in order to improve the spectral coverage
of the data.

Each source was observed typically for 1 min in C, X, U and
K bands, and for 9 min in L band in a single snapshot,
cycling through frequencies. The flux density measurements can
thus be considered as simultaneous.

Two scans were spent on the primary flux density calibrator 3C286,
which was used also for the calibration of the absolute
orientation of the polarization vector. Secondary calibrators were
observed for 1.5 min at each frequency about every 25 min;
they were chosen aiming at minimizing the telescope slewing time.
Accurate positions of the target sources were obtained from the
JVAS catalogue (Patnaik et al. 1992; Browne et al. 1998; Wilkinson
et al. 1998).

The data reduction followed the standard procedures for the VLA,
implemented in the NRAO AIPS software. The L-band imaging has been
quite complicated since a number of confusing sources fall within
the primary beam, and an accurate flux density measurement could
be obtained only once the confusing sources had been cleaned out.
Generally, at least one iteration of phase-only self-calibration
has been performed before the final imaging, although a few
sources required several iterations. Some radio frequency
interferences affected the 1.665 GHz data, so that a few sources
could not be imaged and flux densities could not be derived at
this frequency. A Gaussian fit was performed on the final image by
means of the task JMFIT, and the flux densities of the extended
sources were determined with TVSTAT and IMSTAT.

The L band observations by Dallacasa et al. (2000) were not
adequate to reveal possible extended emission given that 
were taken with the VLA in various configurations and were
too short (1 min). The present deeper exposures detected extended
emission for a considerable fraction of sources (see Sect.~4).

The r.m.s. noise levels in the image plane are relevant only for
flux densities of a few mJy and are generally consistent with the
expected thermal noise. For our bright sources the main
uncertainty comes from the amplitude calibration error,
conservatively estimated to be ($1\sigma$) 3\% for the L, C and X
bands, 5\% for the U band, and 10\% for the K band.

The new multifrequency measurements are listed in columns 6--13 of
Table~\ref{tab:VLA02_flux}, together with the J2000 name (col. 1),
the optical identification and the redshift taken from the NED
database (columns 2 and 3, respectively), the 4.9 GHz flux density
(col. 4) measured by Dallacasa et al. (2000); the labels C and E
in col. 5 stand for \textit{compact} or \textit{extended} sources
respectively, as seen in the L band images (see Sect.~4, Appendix).

The time-lag between the the present data-set and the previous
observations is of about 3 to 4 years. The years of observations
of each source are specified in Table~\ref{tab:peakvalues}.

\begin{table*}
\begin{center}
\caption{Peak frequencies and flux densities}
\begin{tabular}{|c|c|c|l|l|l|l|l|l|l|l|}
\hline
\hline
Name       & & &\multicolumn{2}{c|}{1998}   &  \multicolumn{2}{c|}{1999}& \multicolumn{2}{c|}{2002}& V & $\Delta\nu/(\nu^I_p\cdot \Delta t)  $  \\
\hline
           & & &$S_{p}$ & $\nu_{p}$  &$ S_{p}$ & $\nu_{p}$&$ S_{p}$ & $\nu_{p}$& &  \\
           & & &mJy        & GHz           & mJy        & GHz         & mJy        & GHz         & &   \\
\hline
 0003+2129 &  G* & C  & 272$\pm$5  &  5.7$\pm$0.1  &           &              & 257$\pm$5 & 5.4$\pm$0.1   & 3.02  & 0.018 $\pm$ 0.009 \\
 0005+0524 &  Q  & C  &            &               &235$\pm$6  & 4.13$\pm$0.09& 228$\pm$7 & 3.40$\pm$0.09 & 3.18  & 0.170 $\pm$ 0.03  \\
 0037+0808 &  G? & C  &            &               &287$\pm$5  & 5.9$\pm$0.1  & 287$\pm$6 & 6.2$\pm$0.2   & 0.46  &-0.05  $\pm$ 0.04  \\
 0111+3906 &  G  & E  &            &               &1333$\pm$28& 4.76$\pm$0.06&1303$\pm$28& 4.68$\pm$0.07 & 0.46  & 0.01  $\pm$ 0.01  \\
 0116+2422 &     & C  & 245$\pm$4  & 5.1 $\pm$0.1  &            &             &266$\pm$9  & 6.3$\pm$0.3   & 5.59  &                   \\
 0217+0144 &  Q  & E  &            &               &2557$\pm$105& 18$\pm$3    &            & flat         & 238.27&                \\
 0329+3510 &  Q* & E  & 768$\pm$13 & 6.7$\pm$0.3   &            &             &            & flat         & 21.60 &                \\
 0357+2319 &  Q* & C  & 637$\pm$15 & 12$\pm$1      &            &             &            & flat         & 420.12&                \\
 0428+3259 &  G* & E  & 545$\pm$10 & 7.3$\pm$0.2   &            &             & 539$\pm$11 & 6.8$\pm$0.2  & 1.20  & 0.02 $\pm$ 0.01   \\
 0519+0848 &     & E  & $>$380     & $>$22         &$>$560      & $>$22       & 448$\pm$7  & 7.4$\pm$0.6  & 62.80 &                \\
 0625+4440 & BL  & C  & 575$\pm$17 & 13$\pm$2      &            &             & 237$\pm$4  & 7.4$\pm$1.0  & 160.34&    \\
 0638+5933 &     & C  & 700$\pm$17 & 12$\pm$2      &            &             & 675$\pm$13 & 9.2$\pm$0.7  & 0.83  &                   \\
 0642+6758 &  Q  & C  & 481$\pm$11 & 4.5$\pm$0.1   &            &             & 431$\pm$10 & 4.08$\pm$0.08& 6.77  & 0.10 $\pm$ 0.03   \\
 0646+4451 &  Q  & C  &3371$\pm$118& 15$\pm$2      &            &             &3982$\pm$92 & 11.2$\pm$0.6 & 36.91 & 0.3  $\pm$ 0.1    \\
 0650+6001 &  Q  & C  &1290$\pm$19 & 7.6$\pm$0.3   &            &             & 1143$\pm$27& 5.2$\pm$0.2  & 19.92 & 0.11 $\pm$ 0.01   \\
 1335+4542 &  Q  & C  &737$\pm$18  & 5.1$\pm$0.1   &            &             & 793$\pm$17 & 4.9$\pm$0.1  & 1.23  & 0.03 $\pm$ 0.02   \\
 1335+5844 &     & C  &730$\pm$12  & 6.0$\pm$0.2   &            &             & 736$\pm$12 & 5.5$\pm$0.1  & 1.50  &                   \\
 1407+2827 &  G  & C  &2320$\pm$36 & 5.34$\pm$0.05 &2388$\pm$41 &5.20$\pm$0.09& 2484$\pm$49& 5.01$\pm$0.08& 1.77  & 0.017$\pm$ 0.005  \\
 1412+1334 &     & C  &337$\pm$7   & 4.7$\pm$0.1   &            &             & 345$\pm$8  & 4.18$\pm$0.09& 1.03  &                   \\
 1424+2256 &  Q  & C  &623$\pm$15  & 4.13$\pm$0.07 &            &             & 698$\pm$18 & 3.94$\pm$0.06& 2.11  & 0.05 $\pm$ 0.03   \\
 1430+1043 &  Q  & C  &905$\pm$14  & 6.5$\pm$0.2   &            &             & 887$\pm$39 & 5.7$\pm$0.1  & 4.19  & 0.08 $\pm$ 0.02   \\
 1457+0749 &     & E  &            &               &239$\pm$4   & 5.4$\pm$0.3 & 367$\pm$12 & 1.7$\pm$0.3  & 39.74 &                   \\
 1505+0326 &  Q  & C  &            &               &937$\pm$14  & 7.1$\pm$0.4 & 744$\pm$11 & 6.8$\pm$0.4  & 19.65 & 0.02 $\pm$ 0.04   \\
 1511+0518 & Sy1 & C  &778$\pm$18  & 11.1$\pm$0.4  &            &             & 903$\pm$20 & 10.6$\pm$0.4 & 8.28  & 0.01 $\pm$ 0.01   \\
 1526+6650 &  Q  & C  &427$\pm$13  & 5.7$\pm$0.1   &            &             & 417$\pm$10 & 5.5$\pm$0.1  & 0.53  & 0.04 $\pm$ 0.03   \\
 1603+1105 &     & E  &            &               &277$\pm$4   & 7.7$\pm$0.4 & 229$\pm$9  & 13$\pm$7     & 16.65 &                   \\
 1616+0459 &  Q  & C  &            &               &897$\pm$20  & 4.7$\pm$0.1 & 782$\pm$17 & 4.63$\pm$0.09& 15.20 & 0.02 $\pm$ 0.04   \\
 1623+6624 &  G  & C  &291$\pm$5   & 6.0$\pm$0.2   &            &             & 302$\pm$8  & 6.0$\pm$0.2  & 0.47  & 0.00 $\pm$ 0.01   \\
 1645+6330 &  Q  & C  &628$\pm$18  & 14$\pm$2      &            &             & 629$\pm$17 & 10.1$\pm$0.7 & 2.74  & 0.2  $\pm$ 0.1    \\
 1735+5049 & G?  & C  &972$\pm$18  & 6.4$\pm$0.2   &            &             & 955$\pm$16 & 6.3$\pm$0.3  & 0.15  &                    \\
 1800+3848 &  Q  & E  &            &               &1392$\pm$65 & 17$\pm$3    & 1226$\pm$49& 13$\pm$1     & 1.91  & 0.2  $\pm$ 0.2    \\
 1811+1704 &  Q* & E  &            &               &809$\pm$19  & 12$\pm$1    &            & flat         & 58.74 &                \\
 1840+3900 &  Q  & C  &            &               &201$\pm$3   & 5.7$\pm$0.5 & 169$\pm$4  & 5.2$\pm$0.4  & 20.16 & 0.1  $\pm$ 0.2    \\
 1850+2825 &  Q  & C  &            &               &1567$\pm$32 & 9.1$\pm$0.3 & 1591$\pm$32& 9.5$\pm$0.3  & 4.86  & -0.05$\pm$ 0.06   \\
 1855+3742 &  G* & C  &            &               &426$\pm$42  &4.00$\pm$0.07& 423$\pm$310& 3.81$\pm$0.06& 1.32  & 0.02 $\pm$ 0.01   \\
 2021+0515 &  Q* & C  &            &               &497$\pm$13  &3.75$\pm$0.08& 446$\pm$10 & 4.5$\pm$0.1  & 5.92  &                   \\
 2024+1718 &  S  & E  &            &               &845$\pm$32  &14$\pm$2     & 807$\pm$16 & 8.6$\pm$0.4  & 9.58  & 0.26 $\pm$ 0.06   \\
 2101+0341 &  Q  & E  &            &               &1635$\pm$90 &17$\pm$2     & 737$\pm$212& 3.7$\pm$0.2  & 119.87& 0.53 $\pm$ 0.02   \\
 2123+0535 &  Q  & E  &            &               &2392$\pm$103&18$\pm$4     &            & flat         & 27.49 &                \\
 2203+1007 &  G* & C  &            &               &319$\pm$7   &4.86$\pm$0.07& 312$\pm$7  & 5.0$\pm$0.1  & 5.02  & -0.02$\pm$ 0.02  \\
 2207+1652 &  Q* & E  &            &               &572$\pm$9   &7.4$\pm$0.3  & 252$\pm$7  & 3.5$\pm$0.3  & 228.45&  \\
 2212+2355 &  S  & C  &            &               &1379$\pm$13 &13$\pm$2     &1031$\pm$17 & 9$\pm$1      & 22.62 &  \\
 2257+0243 &  Q  & C  &            &               &            &$>$22        &            & $>$22        & 1.85  &            \\
 2320+0513 &  Q  & E  &            &               &1176$\pm$20 &5.4$\pm$0.2  &            & flat         & 86.24 &            \\
 2330+3348 &  Q  & C  &            &               &558$\pm$9   &5.6$\pm$0.3  &            & flat         & 21.42 &            \\
\hline
\hline
\end{tabular}
\label{tab:peakvalues}
\end{center}
\end{table*}

\section{Spectral analysis}

In order to estimate the peak flux densities, $S_{p}$, and
frequencies, $\nu_{p}$, of the sources, we have fitted the
simultaneous radio spectra at the two epochs using an hyperbolic
function (Dallacasa et al. 2000) of the form:
\begin{equation}
\log(S)= a -\left[b^2+(c\log\nu-d)^2\right]^{1/2} \label{eqhyp}
\end{equation}
with the optically thin and thick spectral indices as asymptotes.
The parameters of this function are related to $S_p$ and $\nu_{p}$
[$\log(\nu_p)=d/c$, $\log(S_{p})=a-b$], so that Eq.~(\ref{eqhyp})
can be rewritten as:
\begin{equation}
\log(S)=\log(S_p)+ b-\left[b^2+c^2(\log\nu-\log\nu_{p})^2\right]^{1/2}.
\label{eqhyp1}
\end{equation}
The best fit values of $\nu_p$ and $S_p$, obtained minimizing the
chi-square function with the Minuit package (CERN libraries), are
reported, with their errors, in Table~\ref{tab:peakvalues}. The
radio spectra of all sources are shown in Fig.~\ref{fig:spectra_1}
where the stars and the filled circles refer to the first and to
the second epoch of simultaneous multifrequency VLA data,
respectively, while the dashed and solid lines indicate the
corresponding fits.

A Kolmogorov-Smirnov (KS) test did not detect any significant
difference among the distributions of the observed turnover
frequencies (shown in Fig.~\ref{fig:peak_twoepochs}) and of the
peak flux densities at the two epochs. This is not surprising
since the time-lag is too short for an evolution of source spectra
(see Sect.~5) to be detected.

The observed peak frequency distributions of candidate HFP
galaxies and quasars are similar (left-hand panel of
Fig.~\ref{fig:peak_gal_stellar}) as a consequence of the
selection criterion, but quasars are found up to much
higher redshifts (Fig.~\ref{fig:dist_redshift}), implying much
higher values of $\nu_p$ in the source frame
(Fig.~\ref{fig:peak_gal_stellar}, right-hand panel). A similar
behavior is found for GPS sources (Snellen 1997; Stanghellini et
al. 1998), whose peak frequencies are, on average, about a factor
of 5 lower.

   \begin{figure*}
   \centering
   \includegraphics{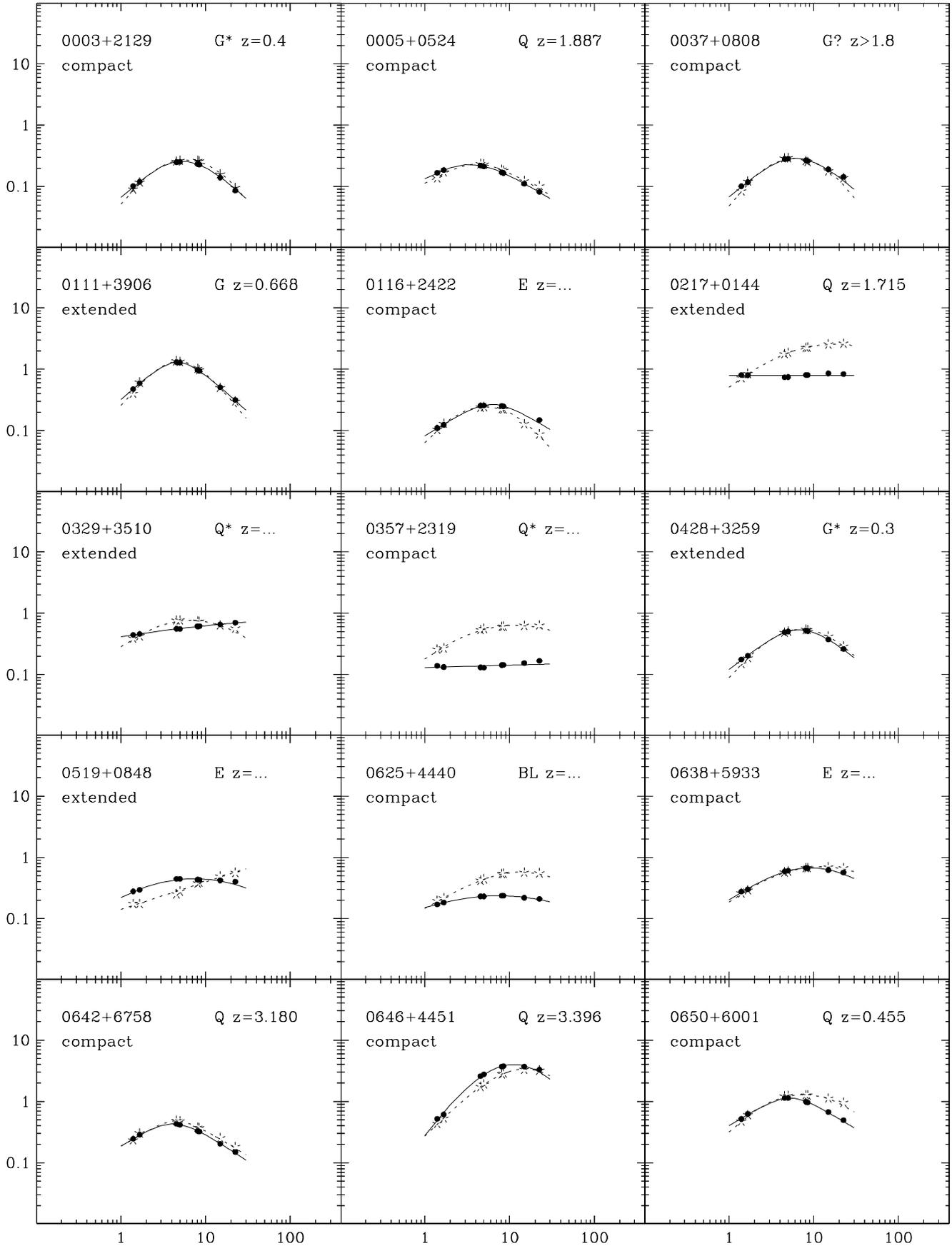}
   \caption{Radio spectra of sources [$S_\nu$ (Jy) vs $\nu$ (GHz)].
   Stars and filled circles represent
   the first and the second epoch of simultaneous multifrequency VLA
   data respectively; the dashed and solid lines show the corresponding
   polynomial fits.}
      \label{fig:spectra_1}
   \end{figure*}

\setcounter{figure}{1}
   \begin{figure*}
   \centering
   \includegraphics{hfp_spc2_square.epsi}
      \caption{(continued)}
      \label{fig:spectra_2}
   \end{figure*}

\setcounter{figure}{1}
   \begin{figure*}
   \centering
   \includegraphics{hfp_spc3_square.epsi}
      \caption{(continued)}
      \label{fig:spectra_3}
   \end{figure*}

\begin{table*}
\begin{center}
\caption[]{Parameters of the extended objects}
\begin{tabular}{c|rr|rr|rr|rr|r}
\hline
\hline
Name & \multicolumn{2}{c|}{$S_{1.4}$ (mJy)} & \multicolumn{2}{c|}{$S_{1.7}$ (mJy)}&\multicolumn{2}{c|}{$S_{\rm C~band}$ (mJy)}&\multicolumn{2}{c|}{$S_{\rm X~band}$ (mJy)} &  LS (arcsec) \\
     & peak & ext & peak& ext& peak&ext  &  peak&ext &      \\
\hline
\hline
J0111+3906 & 477 & 10 & 594 &  7 &     &    &     &  &  17 \\
J0217+0144 & 750 & 75 & 766 & 59 & 736 & 20 & 808 &8 &  $>$6 (4.5 GHz)\\
J0329+3510 & 423 & 80 & 455 & 65 & 553 & 5  &     &  &  35 \\
J0428+3259 & 167 & 12 & 198 &  8 & 492 & 3  &     &  &  $>$8 (4.5 GHz) \\
J0519+0848 & 268 & 17 & 287 & 15 & 444 & 4  &     &  &  10 \\
J1457+0749 & 354 & 13 & 361 & 12 &     &    &     &  &  12 \\
J1603+1105 & 170 & 27 & 175 & 26 & 213 & 2  &     &  &  94 \\
J1800+3848 & 311 &  4 & 329 &  5 &     &    &     &  &  $>$ 6 \\
J1811+1704 & 537 & 15 & 523 & 18 & 488 &  4 & 494 &6 &  22 \\
J2024+1718 & 279 & 12 & 304 & 10 &     &    &     &  &  11 \\
J2101+0341 & 550 & 15 & 609 & 14 &     &    &     &  &  22(EW)$\times$14(NS) \\
J2123+0535 &1949 & 14 &1999 & 34 &     &    &     &  &  11 \\
J2207+1652 & 215 & 50 &     &    & 248 & 12 & 225 & 6&  11 \\
J2320+0513 & 623 & 62 &     &    & 656 & 10 & 721 & 6&   19 (4.5 GHz) \\
\hline
\hline
\end{tabular}
\label{tab:extended_emiss}
\end{center}
\end{table*}

\section{Extended emission}

Within our sub-sample, 14 (31\%) sources show some amount of
extended emission (2 galaxies, 9 quasars and 3 objects without
optical identification) on scales ranging from 6 to 35 arcsec. In
the Appendix we comment on such sources, based on our own
observations and on published VLBA/VLA data. Our images are
presented in Figs.~A1--A10. The flux density of the extended
emission has been measured as the difference between the integrated
flux density over the whole source extension and the
flux density obtained using a point source model (see
Table~\ref{tab:extended_emiss}). Weak emission below the surface
brightness limit ($\sim 0.2$ mJy/beam) of our maps cannot be
excluded.

The total projected linear size of all these sources exceeds the
classical definition of a ``compact'' radio source, i.e. about
15-20 kpc (O'Dea 1998). The total flux density of the
extended emission ranges between a few mJy and 80 mJy, with 1.4
GHz luminosities ranging from $\simeq 10^{24}$ W/Hz (close to the
upper end of the range for FR I sources) to $\simeq 10^{27}$ W/Hz,
well within the luminosity range of FR II sources. The unresolved
to extended luminosity ratios are in the range 4 to 78.

The fraction of sources with extended emission ($\simeq 30\%$) is
larger than in GPS samples, where it is found to be $\simeq 10\%$
(Stanghellini et al. 1990; Baum et al. 1990, Stanghellini et al.,
in preparation). However, no dependence of
the fraction of sources with extended emission on peak frequency 
can be found within our sample: if we split it into two roughly equally
populated sub-samples (with $\nu_p>5.5\,$GHz and $\nu_p<5.5$,
$\nu_p$ being the second epoch peak frequency in the observer's
frame), we have approximately the same fraction of sources with
extended emission in each.

The presence of radio emission on scales of tens of kpc or even
larger can indicate that we are really dealing with an evolved
source but can also be reconciled with the youth scenario under the
hypothesis of recurrent activity proposed by Baum et al. (1990),
whereby a newly born source is propagating amidst the relic of previous
large scale radio activity. At low frequencies (hundreds of MHz)
the extended emission should be dominant, because of its steep
spectrum.

Seven out of the nine quasars with extended emission have strongly
variable spectra. Five of them did not show anymore a convex
spectrum (see Sect.~5) when were re-observed, and therefore are
not HFPs. On the other hand, the two galaxies with some amount of
resolved emission on a kpc scale showed, at both epochs, the
typical peaked spectrum. One is the well known J$0111+3906$
(B$0108+388$) for which Baum et al. (1990) worked out the
recurrent activity hypothesis. We are analyzing high resolution VLBA
observations of J$0428+3259$ to study its pc scale morphology.

\setcounter{figure}{2}
   \begin{figure*}
   \centering
   \includegraphics[width=8 cm]{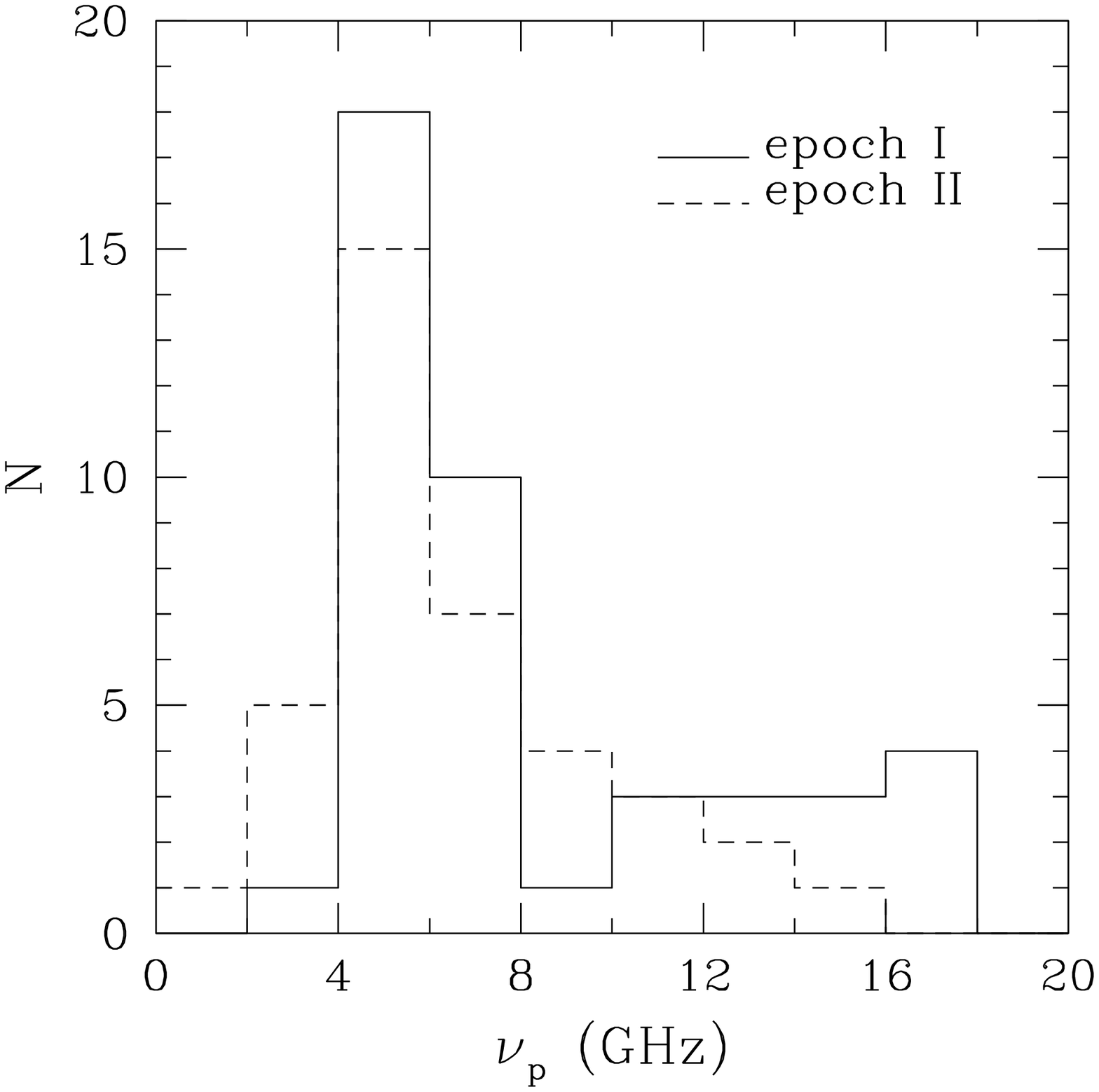}
      \caption{Histogram of observed turnover frequencies
      for the HFP sample at the two epochs.}
      \label{fig:peak_twoepochs}
   \end{figure*}

   \begin{figure*}
   \centering
   \includegraphics[width=8 cm]{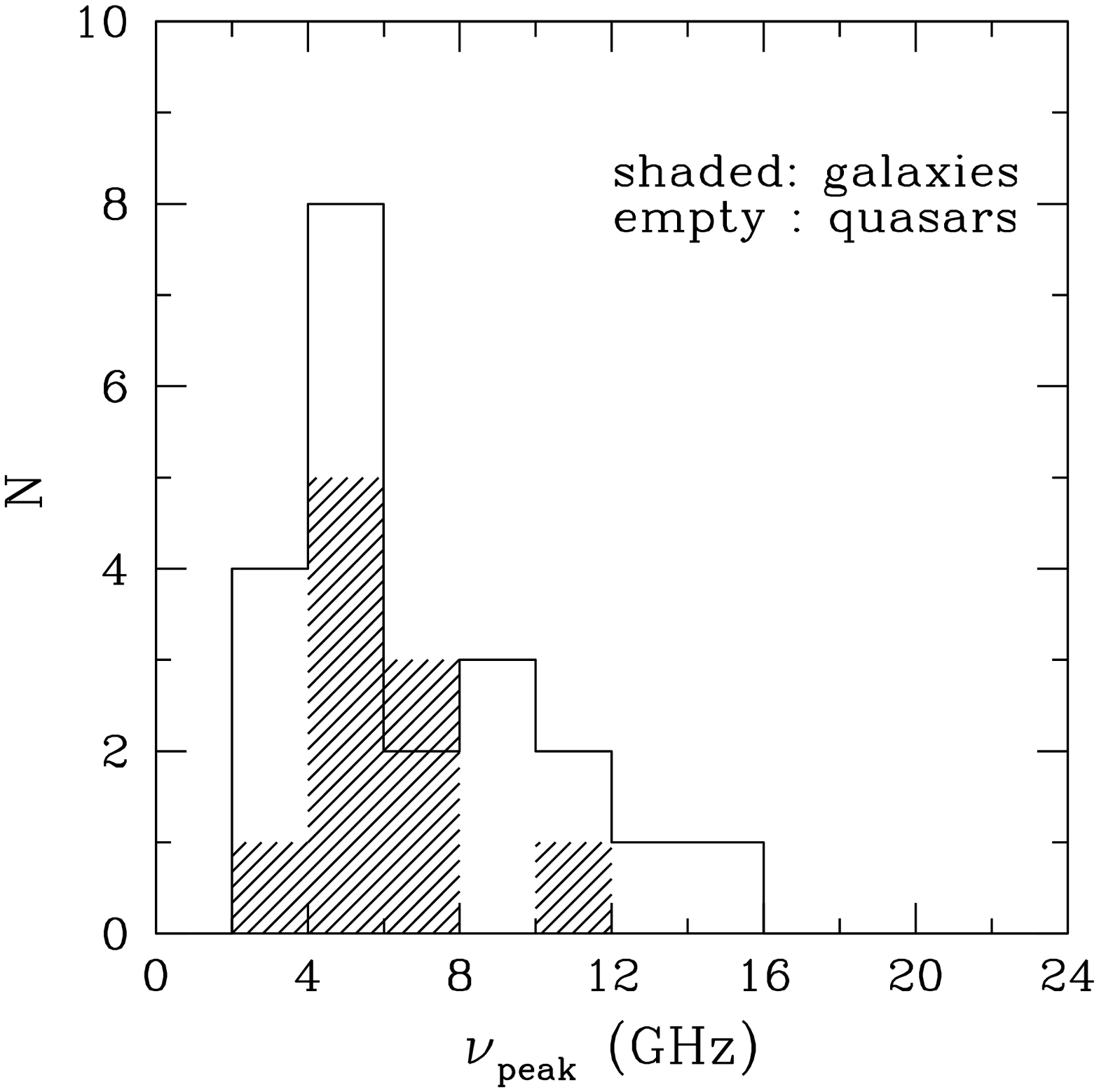}
   \includegraphics[width=8 cm]{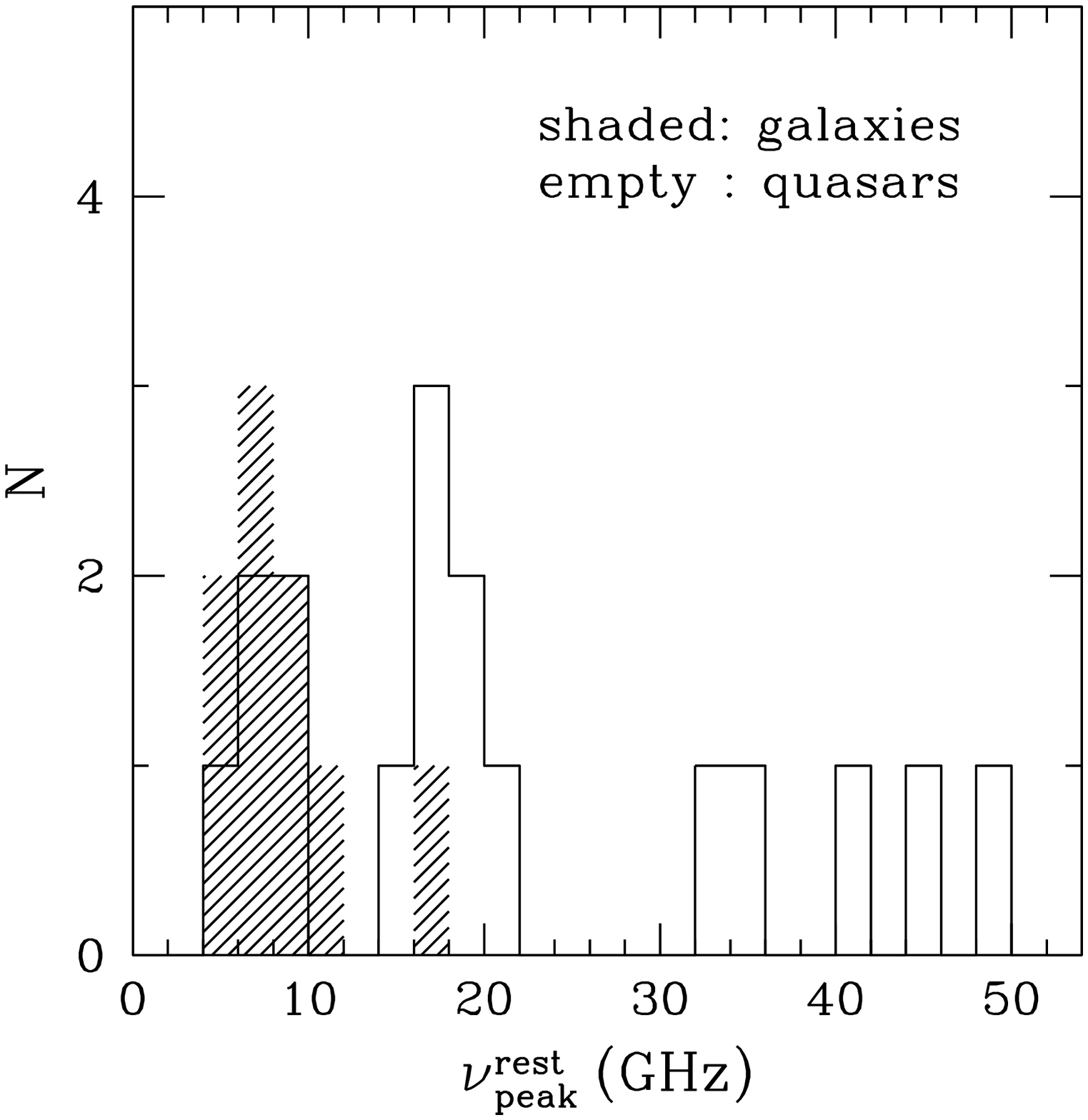}
      \caption{Observed (left) and rest frame (right) turnover frequency
      distributions for galaxies and quasars.}
      \label{fig:peak_gal_stellar}
   \end{figure*}

\section{Analysis of the sample} \label{analysis}

The repeated high sensitivity simultaneous multifrequency
observations separated by 3 to 4 years allow us to
investigate the variability properties of our candidate HFPs. This
analysis may also help discriminating between truly young sources
and flaring blazar components.

The convex form of the spectrum is the first property that should
be preserved in the evolution of young sources. This does not
happen for 7 sources (16\% of the sample), all identified with
quasars (25\% of such objects), whose second epoch spectrum turns
out to be flat. Such sources are labelled as ``flat'' in column 9
of Table~\ref{tab:peakvalues}, and are then classified as blazars.

According to the physical model for the evolution of young
radio sources by Begelman (1996), a newly generated powerful jet
injects energy into the ambient medium inflating a cocoon
consisting of shocked jet material and shocked ambient matter. If
the density of the surrounding medium scales with radius, $r$, as
$\rho\propto r^{-n}$, the linear size, $LS$, of the cocoon
increases with the source age, $t$, as $LS\propto t^\epsilon$,
with $\epsilon=3/(5-n)$. The best determined relationship for
GPS/CSS sources relates the radio turnover frequency to the
projected linear size $\nu_{p}\propto (LS)^{-\delta}$ (O'Dea \&
Baum 1997). It follows that $\nu_{p}$ decreases with time as
$\nu_{p}\propto t^{-\lambda}$, with $\lambda=\delta\cdot\epsilon$.
The expected decrease of $\nu_{p}$ in a time interval $\Delta t$ in 
the source frame (corresponding to $\Delta T = \Delta t(1+z)$ in the observer
frame) is then:
\begin{equation}
\frac{\Delta \nu_p}{\nu_p} = -\lambda\frac{\Delta t}{t}
\label{eq:age}
\end{equation}
The slope of the observed correlation for CSS/GPS sources is
$\delta\simeq 0.65$ but O'Dea \& Baum (1997) show that the
evolution of $\nu_{p}$ with LS may be steeper than the
statistical relation. Assuming Begelman's model, they found  
$\delta\simeq 1.2$ for n=2.

The typical uncertainties associated with our estimates of $\Delta
\nu_p$ are at the several percent level. For $n$ ranging from 0
(uniform density) to 2 (isothermal distribution) and $\Delta T =
3$--4 yr, a clear measurement of the peak frequency decrease due
to the source expansion could be achieved for source ages well below
$100\,$yr. Taking into account that our sources were selected from
the GB87 catalogue, based on observations carried out in
1986--1987, and therefore must have been older than 15--16 yr when
our second set of observations were done, we conclude that sources
with $\Delta \nu_p/[\nu_p \Delta t({\rm yr})]> 0.05$ 
(last column of Table~\ref{tab:peakvalues}) are unlikely
to be HFPs (because too young ages would be implied). We propose
to classify them as blazars, whose observed evolution timescale
are decreased by the Doppler factor. Note that $\Delta \nu_p/
[\nu_p \Delta t({\rm yr})]$ can be computed (and is reported in
Table~\ref{tab:peakvalues}) only for sources
with measured redshift and measured peak frequencies at both epochs.

In principle, an estimate of source ages could be derived from
Eq.~(\ref{eq:age}). In practice, however, given the large
uncertainties on $\Delta \nu_p$ and the additional uncertainties
on $\lambda$ (due to the poor knowledge of the parameter $n$
characterizing the density profile of the medium), only rather
uninteresting lower limits of a few tens of years can be set. On
the other hand, interesting constraints would be provided by more
accurate multifrequency measurements with a 2 or 3 times longer
time-lag.

Although the Kolmogorov-Smirnov (KS) test does not detect any significant
difference in the distribution of turnover frequencies at the two epochs,
it is clear from the last column of Table 2 that most sources have
a smaller $\nu_p$ at the second epoch, consistent with an evolution
towards lower optical depths.

Flux density variability also helps distinguishing between
expanding young objects and aged beamed objects (blazars) whose
radio emission is dominated by a single knot in the jet. While
HFP/GPS sources are thought to expand at mildly relativistic
velocities, the large Doppler factors characterizing blazar jets
boost the variability amplitudes and decrease the corresponding
observed timescales. We have analyzed the variability of our
sources in terms of the quantity
\begin{equation}
V=\frac{1}{m}\sum_{i=1}^{m}\frac{(S_I(i)-S_{II}(i))^2}{\sigma_i^2}~,
\label{eq:V}
\end{equation}
where $S_I(i)$ and $S_{II}(i)$ are the flux densities at the
$i$-th frequency, measured in the first and second epoch,
respectively, $\sigma_i$ is the error on $S_I(i)-S_{II}(i)$, and
$m$ is the number of the sampled frequencies. 

We have noted that the second epoch flux densities at 1.4 GHz are 
systematically higher by about 10\% than the first epoch ones, while the flux
densities at 1.7 GHz at the two epochs appear to match. 
This may be related to an unidentified problem during the 1.4 GHz observation
of the primary calibrator \object{3C286}.
We have therefore chosen to exclude the 1.4 GHz measurements, so that
in the above equation $m=7$.

\begin{figure*}
   \centering
   \includegraphics[width=8 cm]{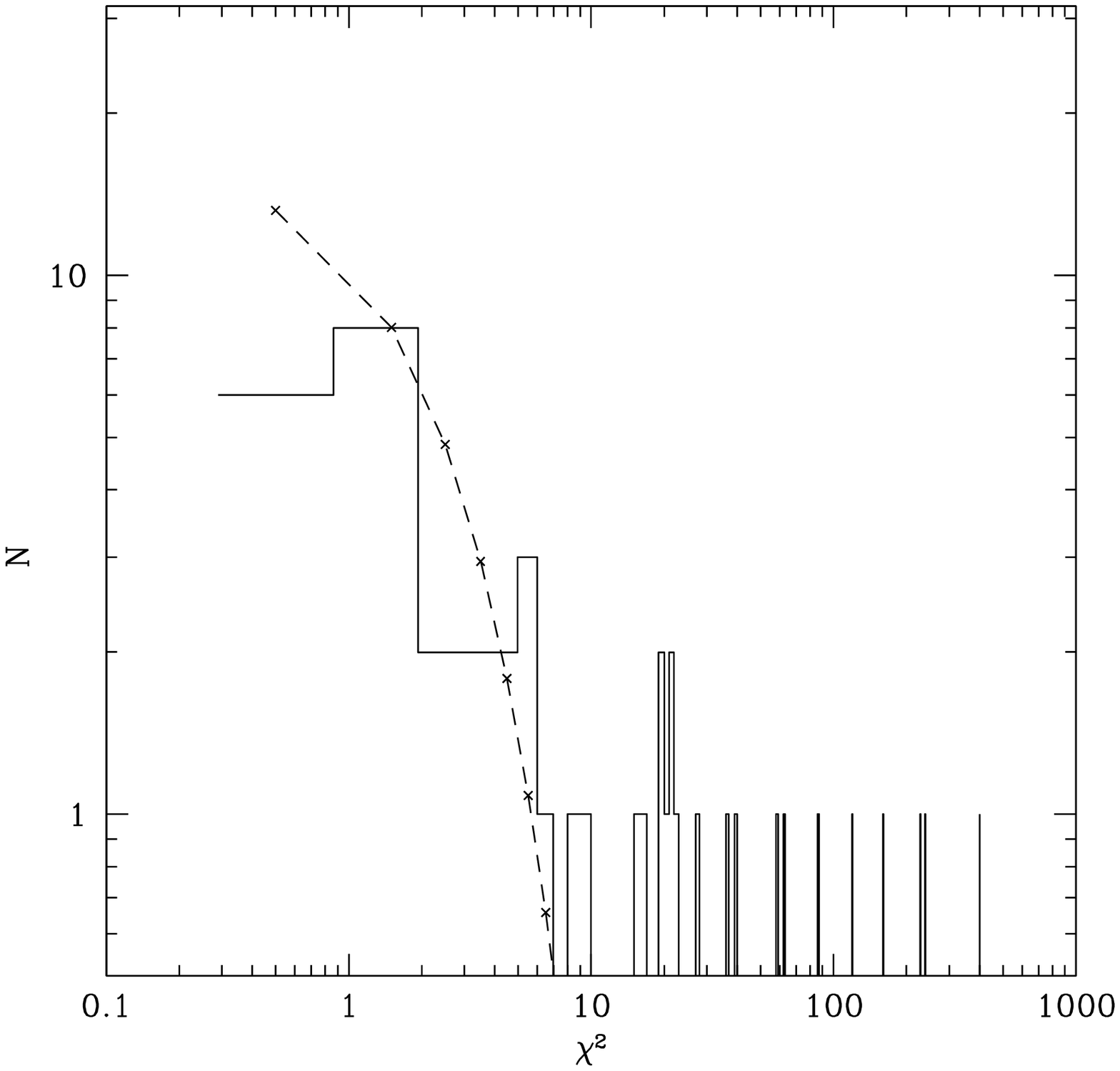}
   \includegraphics[width=8 cm]{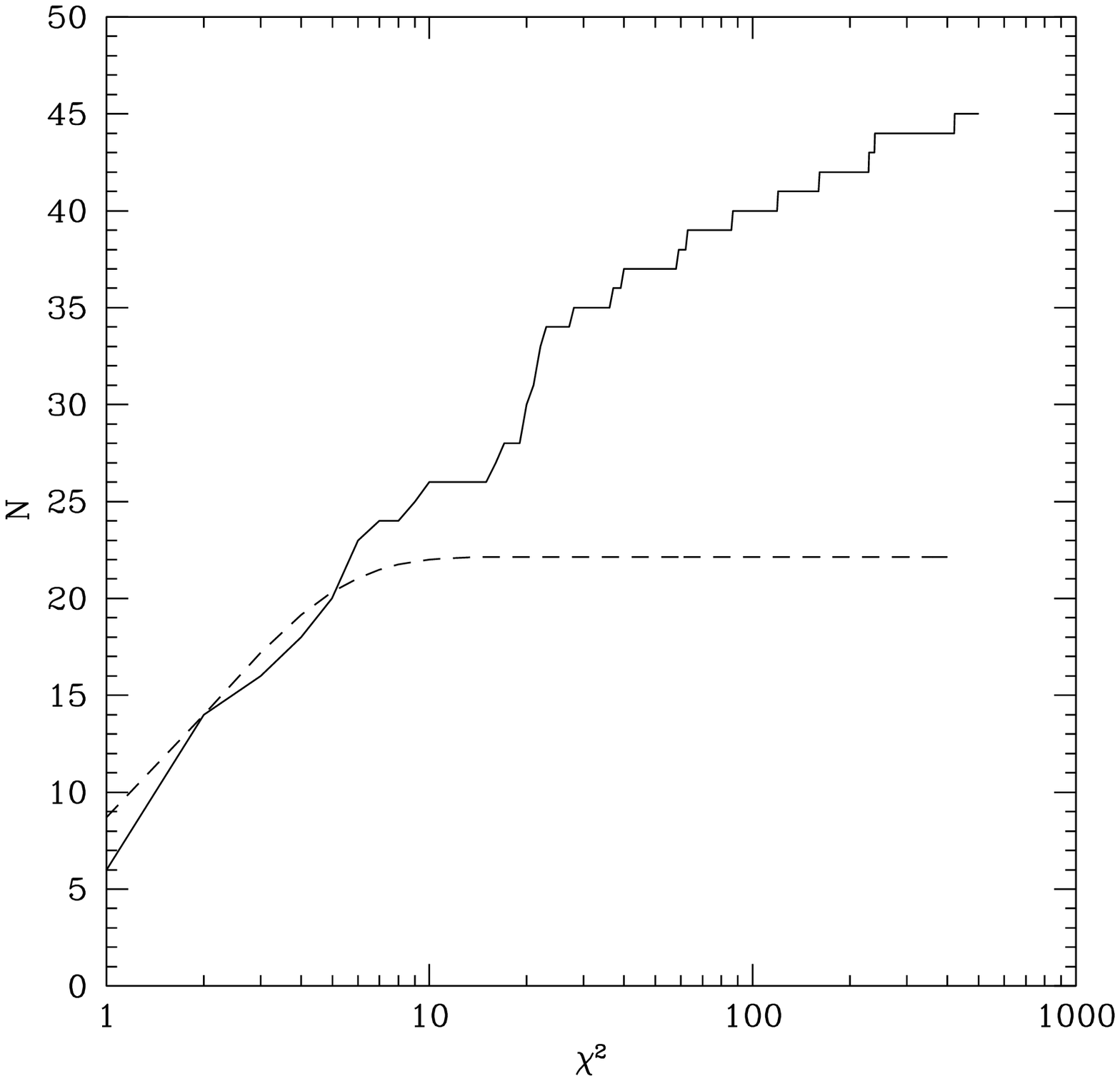}
      \caption{Differential (left) and cumulative (right)
      distribution of $V$, compared with the corresponding $\chi^2$
      distributions for 2 degrees of freedom (dashed lines). The bins of the differential
      distribution correspond to $\Delta \chi^2 =1$.}
      \label{fig:chi_distribution}
   \end{figure*}

For HFP sources, $V$ depends essentially on only two free
parameters, $S_p$ and $\nu_p$ (since the spectral slopes are not
expected to vary), and should therefore have a $\chi^2$
distribution with 2 degrees of freedom.
Figure~\ref{fig:chi_distribution} shows that indeed sources with
low values of $V$ obey such distribution. Since the probability
that a source with $V>9$ is extracted from such distribution is $<
0.01$, we conclude that sources with $V > 9$ are very likely
blazars. On the other hand, blazars may well have small values of
$V$, so that there is no guarantee that sources with $V<9$ are all
truly young. We caution however that these conclusions rely to
some extent on our estimates of errors, dominated by calibration
uncertainties: any significant over- or under-estimate of errors would
impair the match of the distribution of $V$ with the $\chi^2$
distribution with 2 degrees of freedom.

As shown in Table~\ref{tab:peakvalues}, all 10 galaxies,
including the Seyfert 1, have $V<9$, while most quasars (17 out of 28,
including the 7 with flat second-epoch spectra) have $V>9$; of the
7 unidentified sources, 4 have $V<9$. Thus, if we confine
ourselves to sources with $V<9$, we have an almost equal number of
galaxies and quasars, similarly to what is found for GPS samples.
The quasars with $V<9$ have higher median redshift ($z_{\rm
median}\simeq 2.4$) than those with $V>9$, whose median redshift
($z_{\rm median}\simeq 1.7$) is closer to that of flat spectrum
radio quasars in the Parkes quarter-Jy sample ($z_{\rm
median}\simeq 1.4$). 

An alternative approach may be to consider as blazar candidates 
all sources with variability significant at more than $3\sigma$. 
In this case the boundary between candidate HFP and likely blazars has to be
set at $V=3$ (as appropriate for a $\chi^2$ distribution with 1 degree of
freedom) rather than at $V=9$. If so, 9 more sources (3 galaxies,
including the type 1 Seyfert, 5 quasars and 1 unidentified source)
add to the list of candidate blazars.

Another interesting issue is the relationship between peak
frequencies and peak luminosities, $L_p$. Opposite trends are predicted
by current models for GPS sources: while Begelman's (1996) model
implies a decrease of $L_p$ with decreasing $\nu_p$, Snellen et
al. (2000) predict an increase during the HFP/GPS phase of the
source evolution. In both cases, however, the dependence of $L_p$
on $\nu_p$ is weak, and therefore easily swamped by variances in
both quantities. In the case of relativistic beaming with Doppler
factor $\delta$, the observed peak luminosities scale as
$\delta^3$, while the observed peak frequencies scale as $\delta$,
so that sources with equal intrinsic $L_p$ and $\nu_p$, but
different $\delta$'s are observed to have $L_p \propto \nu_p^3$;
this relation is much steeper than the $L_p$--$\nu_p$ relation
predicted by both Begelman's and Snellen's models.

\begin{figure*}
   \centering
    \includegraphics[width=8 cm]{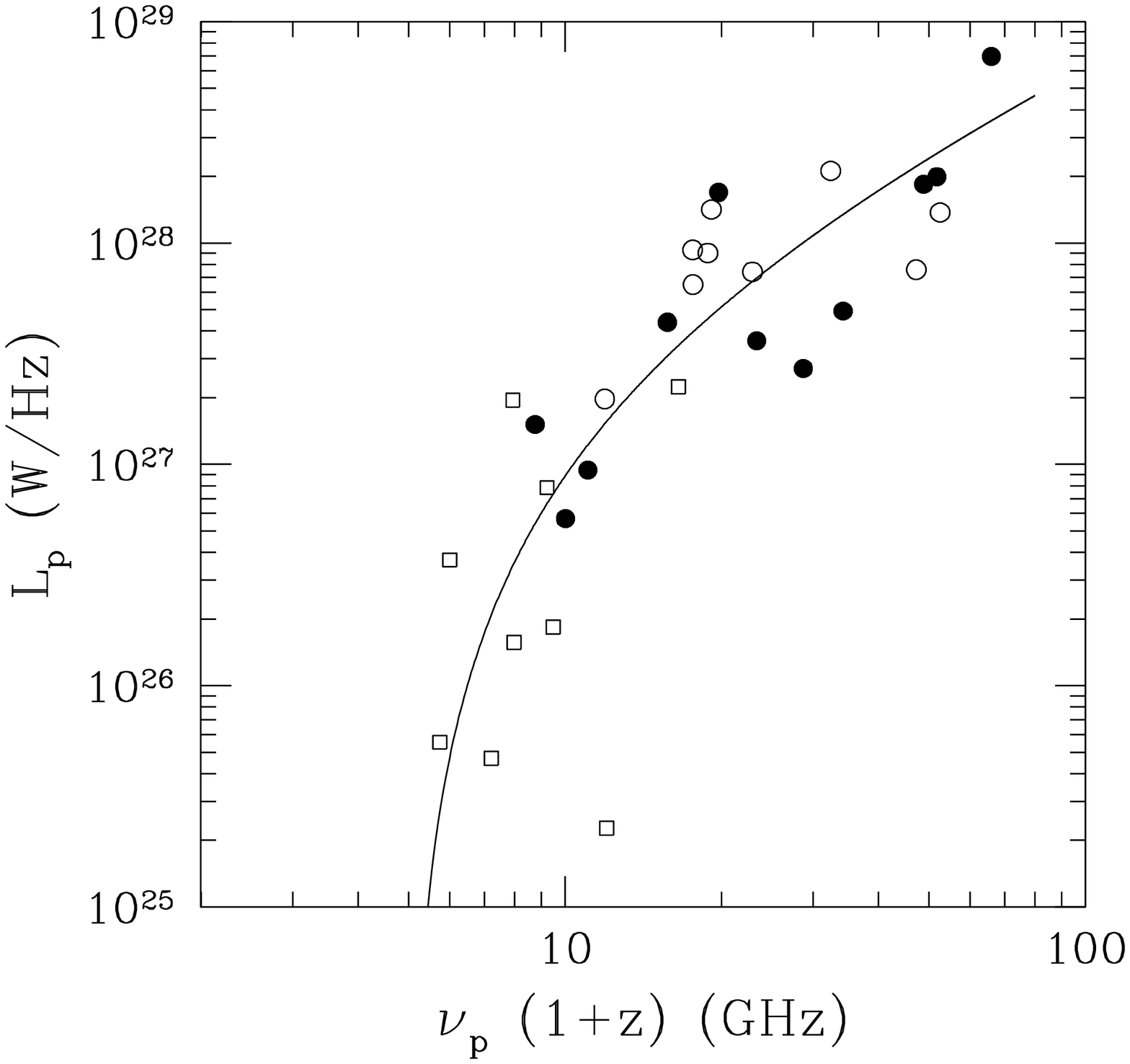}
       \caption{Peak luminosity versus peak frequency
       for all objects with measured redshift.
       Circles: quasars, squares: galaxies. Open symbols refer to
       sources with $V\le9$, filled symbols to sources with $V>9$.
       The solid line shows the distribution in the $L_p$--$\nu_p$ plane of sources with equal
observed $S_p=300\,$mJy and $\nu_p=5\,$GHz at different redshifts.  }
       \label{fig:Lpnup_all}
   \end{figure*}

We have investigated this relationship for all objects (see
Fig.~\ref{fig:Lpnup_all}), using first epoch data, when all
sources but two had a well defined spectral peak. Although a clear
correlation seems to be present, we must beware of the effect of
the redshift distribution: as an example, the solid line in
Fig.~\ref{fig:Lpnup_all} shows the distribution in the
$L_p$--$\nu_p$ plane of sources with equal observed $S_p=300\,$mJy
and $\nu_p=5\,$GHz at different redshifts. Clearly the shape of
the correlation is not far from that induced by the effect of
redshift. Nevertheless, Kendall's partial correlation
coefficient indicates a statistically significant positive
correlation between $L_p$ and $\nu_p$ after the influence of
redshift has been eliminated. The probability of no correlation is
0.44\% if we consider only stellar objects with $V>9$ (which are
probably blazars; we have 12 such objects with redshift, but for
one of them the peak frequency is undefined). For stellar objects
with $V\le9$ the probability of no correlation increases to 6\%
so that the correlation is only marginally significant. For
galaxies there is no indication of a significant correlation
(probability of no correlation 38\%).

Models for GPS/HFP sources also predict a correlation between peak
luminosity and linear size, whose slope is related to the slope
$n$ of the density profile of the surrounding medium.
Unfortunately, our VLA observations provide only upper limits to
the source sizes, since sources are unresolved. A search of
databases available in the VLBA Web site has yielded maps at
various frequencies for 15 out of the 18 sources with $V\le9$ and
known redshift. Of these, 8 were resolved and for them we
estimated the angular size. For the remaining 10 we adopted upper
limits corresponding to the beam size. One source has not been
included in the analysis because it has not a defined spectral
peak at the first epoch of VLA observations. These data show no
indication of a correlation between the linear size and the peak
luminosity (probability of no correlation $\simeq 0.7$) and do not
confirm, for our sample, the correlation between peak flux density
and angular size (probability of no correlation $\simeq 0.6$)
reported by Snellen et al. (2000) for their GPS sample.

\section{Discussion and conclusions}

The spectral selection adopted by Dallacasa et al. (2000) is
independent of optical identification and redshift. As a result,
both galaxies and star-like objects (``quasars'') are present in
the original sample of bright HFP candidates. Quasars are almost
three times more numerous than galaxies, consistent with the
previously noted decrease of the galaxy to quasar ratio in GPS
samples selected at increasing turnover frequency (Fanti et al.
1990; O'Dea 1998; Stanghellini et al. 1998, 2003).

However, our new multifrequency observations, with a time lag of 3
to 4 years, indicate that the sample of candidate HFP quasars is
likely to be strongly contaminated by beamed objects, with
variability properties consistent with those of blazars.

Seven out of the 28 star-like sources (25\%) no longer have a
peaked spectrum, and 5 of them have extended emission. Such
sources are therefore classified as blazars, caught by Dallacasa
et al. (2000) during a flaring phase of a strongly self-absorbed
component.

Seven additional star-like sources show a variation of the peak
frequency, significant at a $>3\sigma$ level, larger than expected
if HFP/GPS sources expand at mildly relativistic velocities
($\Delta \nu_p/[\nu_p \Delta t({\rm yr})]> 0.05$), suggesting the
presence of relativistic beaming effects. The only galaxy
(1407+2827) with a
peak frequency variation significant at $>3\sigma$ level has $\Delta
\nu_p/[\nu_p \Delta t({\rm yr})]= 0.017$.

Another criterion to discriminate between truly young sources
and blazar is the variability index $V$ [Eq. (\ref{eq:V})]. The
distribution of $V$ for the less variable sources is close to the
$\chi^2$ distribution with two degrees of freedom, as may be
expected since variations of spectra of young sources are
controlled by two parameters, $S_p$ and $\nu_p$. If so, sources
with $V>9$ have a probability $<0.01$ of being extracted from such
a distribution and are therefore likely blazars. We have 20
sources with $V>9$, 17 of which are star-like (including the 7
with flat spectrum at the second epoch and 4 of the 7 with $\Delta
\nu_p/[\nu_p \Delta t({\rm yr})]> 0.05$) and 3 unidentified
sources. The median r.m.s. variations for these sources are
$\simeq 30\%$ ($27\%$ for quasars only), close to those found for
blazars ($\simeq 32\%$; Ciaramella et al. 2004).

The variability timescales in the frequency range 5--15 GHz of bright blazars, 
derived from their structure function, are $\simeq 2\,$yr in the source frame
(Hughes et al. 1992; Ciaramella et al. 2004). 
To convert the time-lag between the two sets of
observations into the mean time interval in the source frame, we
need to divide it by $(1+z_{\rm mean})$.
Only 11 of the 17 starlike sources with $V>9$ have
measured redshift, with a median value of 1.7. We may thus expect
that $\sim 50\%$ of blazars flaring at the moment of the first set
of observations have recovered their baseline flat spectrum by the
epoch when they where re-observed. For comparison, 7 out of 17
(i.e. $\simeq 41\%$) were indeed found to have a flat
spectrum in the second observing run.

There is evidence of a positive correlation between $V$ and
$\nu_p$ (probability of no correlation $\sim 0.01$), indicating
that the fraction of blazars is higher among objects with higher
peak frequency. This is not surprising because young objects
become increasingly rare since their lifetimes decrease with
increasing $\nu_p$.

We have found a statistically significant positive correlation
between the peak luminosity and rest-frame peak frequency of
quasars, after having removed the effect of the redshift
distribution (Kendall's partial correlation test). A relationship
among these two quantities is expected both in the case of GPS/HFP
sources and of blazars, but with a much steeper slope for the
latter sources. While previous studies enlightened a
$S_{p}$--$\nu_{p}$ correlation for a few individual blazars (e.g.,
Stevens et al. 1996) our analysis indicates that it holds for
flaring blazars as a class.

There are thus various pieces of evidence, albeit circumstantial,
that most quasar HFP candidates are actually objects where the
effects of beaming are relevant (blazar like) even if
this does not necessarily require that we are seeing evolved/old
objects.

While sources with $V>9$ are likely blazars there is no guarantee
that those with $V<9$ are truly young sources. Some of them may
really be blazars in a relatively quiescent phase. The 3 such
objects with $\Delta \nu_p/[\nu_p \Delta t({\rm yr})]> 0.05$ may
belong to this category.

As for galaxies, all 10 HFP candidates have $V<9$.
Their peak luminosities and rest-frame peak frequencies
do not show evidence
of a correlation, not surprisingly given the smaller range spanned
by both quantities (compared to the case of quasars) and the
weaker dependence of $L_p$ on $\nu_p$ expected for GPS/HFP
sources, easily swamped by intrinsic dispersions of both
quantities.

Two galaxies show extended emission compatible with recurrent
activity. According to Begelman's model the synchrotron emission
decreases with source age $t$ as $L\propto t^{-\eta}$ with
$\eta=(n+4)/[4(5-n)]$. In the case of recurrent activity, the HFP
to extended emission age ratio can then be estimated as $t_{\rm
HFP}/t_{\rm ext}=(L_{\rm ext}/L_{\rm HFP})^{1/\eta}$,
where $L_{\rm ext}$ and $L_{\rm HFP}$ are the total {\it
emitted} radio luminosities, that can be taken as proportional to
the {\it emitted} flux densities at 1.4 GHz. In the case of the
HFP component, the latter can be estimated extrapolating to 1.4
GHz the peak flux density. The age ratio derived from Begelman's
model depends on the slope $n$ of the density profile of the
surrounding medium. Adopting $n=2$ we get, for J$0111+3906$,
$t_{\rm ext}/t_{\rm HFP}\simeq 9\times 10^4$, and for
J$0428+3259$ $t_{\rm ext}/t_{\rm HFP}\simeq
2\times10^4$, consistent with the intermittency timescales
suggested by Reynolds \& Begelman (1997). Although any conclusion
at this stage is premature, this illustrates the potential of
deeper low frequency observations to shed light on the nature of
radio activity.

In conclusion, while the variability properties of candidate HFP
galaxies may be consistent with expectations for truly young,
possibly recurrent, sources, those of most quasars are consistent
with those of blazars. If we set the boundary between HFP
candidates and blazars at $V=3$ (see Sect.~\ref{analysis}),
 3 galaxies would host a blazar nucleus,
and 22 out of the 28 starlike sources and 4 out of the 7
unidentified sources would be blazar candidates. The median
variability amplitude of sources with $V>3$ is $\simeq 22\%$, 
slightly lower than, but still not far from that found for blazar
samples. Milli-arcsec morphology at different frequencies,
polarization measurements and possibly the study of the spectral
index distributions in the optically thin regime can help
distinguishing between flaring blazars and truly young objects.

High resolution VLBA observations at two frequencies in the
optically thin part of the spectrum (between 8.4 and 43.2 GHz) of
50 objects of the original sample have already been acquired. The
analysis of the data is in progress.

\begin{acknowledgements}
We thank the referee, Prof. Roberto Fanti, for useful comments
that helped improving substantially the manuscript.
The VLA is operated by the U.S. National Radio Astronomy
Observatory which is a facility of the National Science Foundation
operated under a cooperative agreement by Associated Universities,
Inc. This work has made use of the NASA/IPAC Extragalactic
Database NED which is operated by the JPL, California Institute of
Technology, under contract with the National Aeronautics and Space
Administration. The authors acknowledge financial support from the
Italian MIUR under grant COFIN-2002-02-8118.
\end{acknowledgements}

\appendix
\section{notes on the individual sources}
\begin{figure*}
   \centering
   \includegraphics[width=15.5cm]{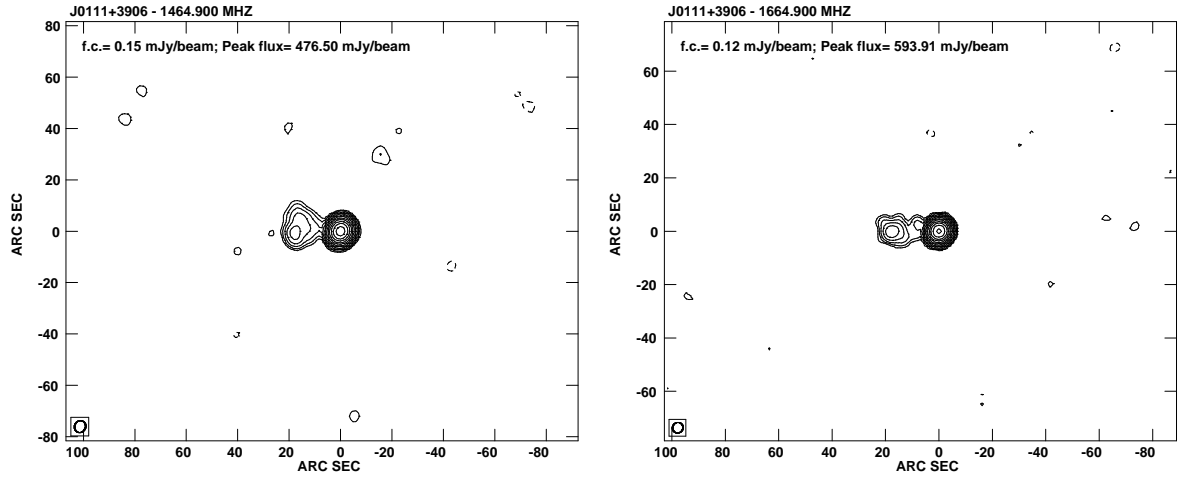}
       \caption{J0111+3906; the restoring beam is in the Bottom Left Corner of each figure, the conturn levels are
$\pm 1$, 2, 4, 8, 16, 32, 64, 128, 256, 512, 1024, 2048, 4096 times the first conturn (f.c.) corresponding to
3 times the r.m.s. noise of the image.}
       \label{fig:J0111+3906}
   \end{figure*}

\begin{figure*}
   \centering
   \includegraphics[width=15.8cm]{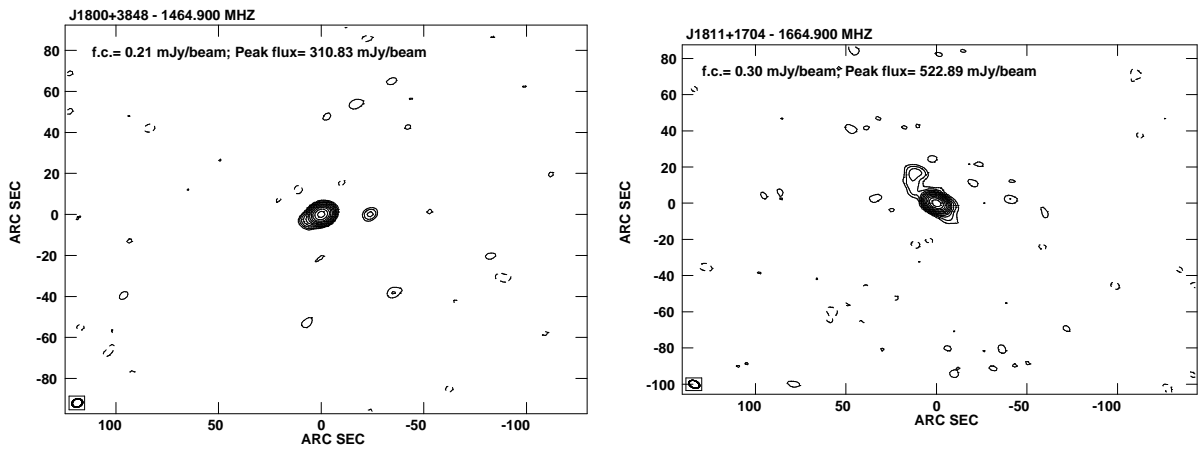}
       \caption{J1800+3848 (left) at 1.465 GHz and J1811+1704 (right) at 1.665 GHz}
       \label{fig:J1800+3848}
   \end{figure*}

\begin{figure*}
   \centering
   \includegraphics[width=15.5cm]{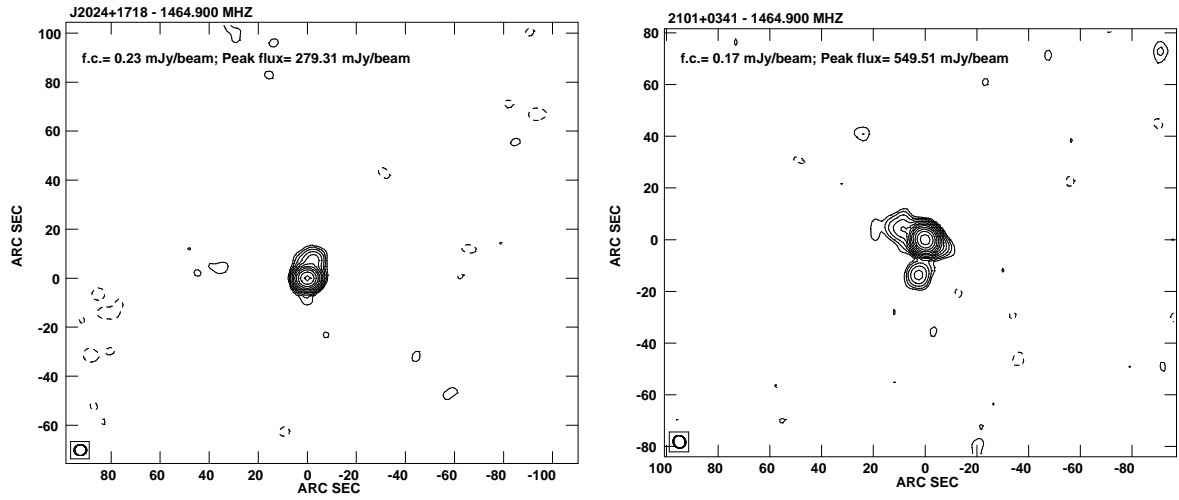}
       \caption{J2024+1718 (left) at 1.465 GHz and J2101+0341 (right) at 1.465 GHz}
       \label{fig:J2024+1718}
   \end{figure*}

{\bf{J0111+3906}}: well known galaxy at $z=0.668$. We have
detected extended emission on kpc scale (see Fig.~\ref{fig:J0111+3906}),
already discovered and imaged by Baum et
al. (1990), Taylor et al. (1996), Stanghellini (2003), at
different resolutions. The extended emission is in contrast with a
recent origin for the radio activity in this source
but can be explained if the radio source is recurrent 
(Baum et al. 1990).\\
{\bf{J0217+0144}}: quasar at $z=1.715$. The source is slightly
resolved in the NW-SE direction in the L band. At higher
resolution, in the C and X bands, two opposite regions of emission
with respect to the central compact component
 are detected (see Fig.~\ref{fig:J0217+0144}).\\
{\bf{J0329+3510}}: stellar object. We see structure on arcsecond
scale (lobes-hotspots). The extended emission is still detected,
even if very weak, in the C band (see Fig.~\ref{fig:J0329+3510});
a jet like feature is emerging from the
core towards the brightest lobe. \\
{\bf{J0428+3259}}: galaxy with estimated redshift $z=0.3$
(Dallacasa et al. 2002), based on the Hubble diagram in Snellen et
al. (1996). The source is slightly resolved along the NW-SE axis
at both 1.465 GHz and 1.665 GHz (Fig.~\ref{fig:J0428+3259}), but
no resolved emission is found at 5 GHz, due to sensitivity limitations.\\
{\bf{J0519+0848}}: optically unidentified. It is slightly resolved
with a hint of weak emission in the NW direction
(see Fig.~\ref{fig:J0519+0848}).\\
{\bf{J1457+0749}}: optically unidentified. It is slightly resolved
in the NW direction (see Fig.~\ref{fig:J1457+0749}). \\
{\bf{J1603+1105}}: optically unidentified. The NE
component can be either associated with the central object or can
be an unrelated source (see Fig.~\ref{fig:J1603+1105}); according
to Snellen (1998) there is a 6\% chance of finding an unrelated
NVSS radio source with a flux density of $> 5\,$mJy within a radius of 100 arcsec
from the target source.\\
{\bf{J1800+3848}}: quasar at $z=2.092$. At 1.465 and 1.665 GHz the
source appears slightly resolved but the estimated flux density of
the extended emission
is of a few mJy only (see Fig.~\ref{fig:J1800+3848}). \\
{\bf{J1811+1704}}: stellar object. A very weak component at
NW is clearly visible both at 1.465 GHz
and 1.665 GHz (see Fig.~\ref{fig:J1800+3848}).\\
{\bf{J2024+1718}}: stellar object at $z=1.050$. The emission is
extending from the core
to the North for 11 arcsec (see Fig.~\ref{fig:J2024+1718}).\\
{\bf{J2101+0341}}: quasar at $z=1.013$. This source has a very
complex structure. Two components are aligned with the NS
axis for a total angular size of 14 arcsec. Furthermore there is
a diffuse low-brightness emission in the NE direction with
angular size of 22 arcsec (see Fig.~\ref{fig:J2024+1718}).\\
{\bf{J2123+0535}}: quasar at $z=1.878$. There is a very weak
emission toward the South of the strong compact component (see
Fig.~\ref{fig:J2123+0535}).\\
{\bf{J2207+1652}}: stellar object. On arcsec scales it shows two
components, the southern one being weaker and smaller. This
component
is still clearly visible in the X band (see Fig.~\ref{fig:J2207+1652}).\\
{\bf{J2320+0513}}: quasar at $z=0.622$. Extended emission is
revealed in both the L and the C band.
The maximum angular size in the NW-SE direction
is of 19 arcsec. In the C band
the extended emission has two irregular components on
both sides of the unresolved nucleus, the southern one being weaker
than the northern one (see Fig.~\ref{fig:J2320+0513}).

   \begin{figure*}
   \centering
   \includegraphics[width=15.5cm]{0217ext_new.epsi}
       \caption{J0217+0144}
       \label{fig:J0217+0144}
   \end{figure*}

   \begin{figure*}
   \centering
   \includegraphics[width=15.5cm]{0329ext_new.epsi}
       \caption{J0329+3510}
       \label{fig:J0329+3510}
   \end{figure*}

   \begin{figure*}
   \centering
   \includegraphics[width=15.5cm]{0428ext_new.epsi}
       \caption{J0428+3259}
       \label{fig:J0428+3259}
   \end{figure*}

   \begin{figure*}
   \centering
   \includegraphics[width=15.5cm]{0519ext_new.epsi}
       \caption{J0519+0848}
       \label{fig:J0519+0848}
   \end{figure*}

   \begin{figure*}
   \centering
   \includegraphics[width=15.5cm]{1457ext_new.epsi}
       \caption{J1457+0749}
       \label{fig:J1457+0749}
   \end{figure*}

   \begin{figure*}
   \centering
   \includegraphics[width=15.5cm]{1603ext_new.epsi}
       \caption{J1603+1105}
       \label{fig:J1603+1105}
   \end{figure*}

   \begin{figure*}
   \centering
   \includegraphics[width=6.5cm]{2123ext_new.epsi}
       \caption{J2123+0535}
       \label{fig:J2123+0535}
   \end{figure*}

   \begin{figure*}
   \centering
   \includegraphics[width=15.3cm]{2207ext_new.epsi}
       \caption{J2207+1652}
       \label{fig:J2207+1652}
   \end{figure*}

   \begin{figure*}
   \centering
   \includegraphics[width=15.3cm]{2320ext_new.epsi}
       \caption{J2320+0513}
       \label{fig:J2320+0513}
   \end{figure*}

\end{document}